\documentclass[prl,twocolumn]{revtex4}
\usepackage{graphicx}
\usepackage{dcolumn}
\usepackage{bm}
\usepackage{amsmath}
\usepackage{amssymb}
\usepackage{mathrsfs}
\usepackage{amsfonts}

\begin{document} 

\title{Superradiant phase transitions with three-level systems}

\author{Alexandre Baksic}
\author{Pierre Nataf}
\author{Cristiano Ciuti}
\affiliation{Laboratoire Mat\'eriaux et Ph\'enom\`enes Quantiques,
Universit\'e Paris Diderot-Paris 7 and CNRS, \\ B\^atiment Condorcet, 10 rue
Alice Domon et L\'eonie Duquet, 75205 Paris Cedex 13, France}

\begin{abstract}
We determine the phase diagram of $N$ identical three-level systems interacting with a single photonic mode in the thermodynamical limit ($N \to \infty$) by accounting for 
the so-called diamagnetic term and the inequalities imposed by the Thomas-Reich-Kuhn (TRK) oscillator strength sum rule.
The key role of transitions between excited  levels and the occurrence of first-order phase transitions is discussed.
We show that, in contrast to two-level systems, in the three-level case the TRK inequalities do not always prevent a superradiant phase transition in presence of a diamagnetic term. 

\end{abstract}

\maketitle

The collective superradiant coupling of a large number of atoms (or artificial atoms) has attracted a considerable interest since the pioneering work of Dicke\cite{Dicke} and is now
the focus of many recent studies in cavity\cite{Dimer,esslinger,Nagy,Baumann,Bhaseen} and circuit\cite{NatafPRL1,NatafNAT,NatafPRL2} quantum electrodynamics.
The well-known Dicke model describes the coupling between a collection of two-level systems and a single photon mode. Remarkably, for increasing light-matter
coupling such a model predicts a superradiant phase transition \cite{Lieb,Carmichael,Brandes}, with a doubly degenerate ground state above a critical vacuum Rabi coupling. 
The so-called superradiant phase is characterized by a spontaneous polarization of the atoms and a spontaneous coherence of the cavity field in the ground state.
In the case of  time-independent Hamiltonians, photons in the ground state are 'virtual' (i.e., bound in the cavity) and they cannot be radiated out of the cavity\cite{CiutiPRA2006} unless some non-adiabatic modulation of the Hamiltonian is applied\cite{DeLiberatoPRL2007} (in analogy with the dynamical Casimir effect\cite{Casimir}). In this sense, the term 'superradiant' currently used for the Hepp-Lieb\cite{Lieb} ground state is unfortunate, because it was originally introduced by Dicke\cite{Dicke} for collective excited state radiative decay and not for ground state properties.
Even in spite of no extracavity emission from the ground state, occurrence of a superradiant critical point can be in principle monitored by measuring the dispersion of the collective excitations via standard optical techniques (e.g., through transmission spectroscopy).  In the case of non-equilibrium phase transitions for pumped open systems, there is not a true ground state\cite{Dimer,esslinger,Nagy,Baumann,Bhaseen}, but instead a stationary state, which is accompanied by the emission of real photons.  

However, for the case of electric dipole transitions, the Dicke model does not include the so-called diamagnetic term, which is proportional to the squared electromagnetic vector potential present in the minimal 
coupling Hamiltonian describing light-matter interaction in the non-relativistic regime.
It is known that in the case of two-level real atoms, such a diamagnetic term is crucial, because it forbids the phase transition as a result of the TRK oscillator strength sum rule\cite{pol1,NatafNAT}.
We point out that such no-go theorem cannot be necessarily applied to time-dependent Hamiltonians with applied dressing fields\cite{esslinger}, to the case of magnetic dipole coupling\cite{Knight} and to more complex effective systems simulating the Dicke Hamiltonian with different degrees of freedom\cite{esslinger,Nagy}. Moreover, the no-go theorem\cite{pol1,NatafNAT} is formulated for two-level systems, while the study of the multilevel case has been initiated only recently\cite{Viehmann,Ciuti,Hayn}

A generalized model including three-level atoms in the lambda configuration coupled to two photon modes has been theoretically investigated in a recent paper\cite{Hayn}  showing a very  rich phase diagram
with superradiant transitions of both second and first order (in the case of two-level atoms the superradiant transition is of second order).  In such an interesting work\cite{Hayn}, the diamagnetic term has not been included and the particular two-color lambda configuration has been considered, so the general case of an arbitrary three-level system needs to be explored.
In a recent letter\cite{Viehmann}, it has been claimed that for multilevel atoms coupled to a single photon mode the no-go theorem can be generalized using again the TRK oscillator strength sum rule: in the proof reported in such a work\cite{Viehmann}, through a perturbative argument, it is assumed that transitions between excited states can be always neglected in the thermodynamical limit: this is rather surprising\cite{Ciuti}, since in  the work by Hayn et al.\cite{Hayn}, transitions between excited states play instead a crucial role in the thermodynamical limit and are responsible for the first-order transition boundaries. It is therefore a fundamental problem to explore superradiant phase transitions with arbitrary three-level systems.

In this letter, we investigate the existence of superradiant phase transitions of a system consisting of $N$ three-level systems coupled to a single photon mode including the diamagnetic term in the Hamiltonian and TRK inequalities.
We show the rich phase diagram in the thermodynamical limit ($N \to \infty$), obtained via a multilevel Holstein-Primakoff approach.
We have studied particular configurations (ladder, lambda, V-type) and also the general three-level coupling configuration.
We find that for three-level systems TRK inequalities do not always prevent superradiant phase transitions and that excited state transitions play a crucial role in the thermodynamical limit.
\begin{figure}[t!]
   \begin{center}
   \includegraphics[width=150pt,angle=90]{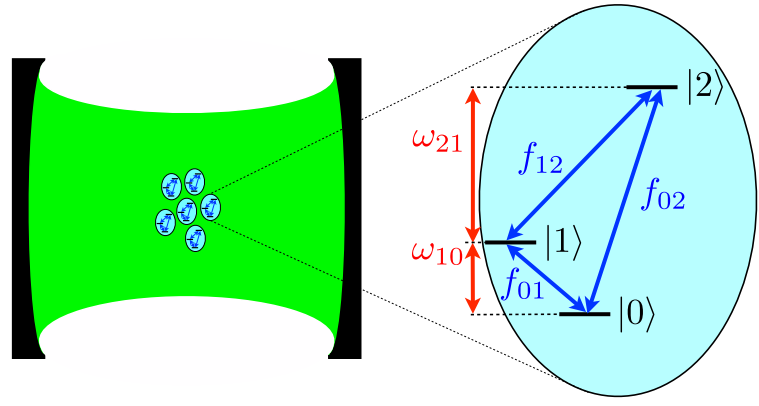}
   \end{center}
   \caption{A sketch of the considered system consisting of $N$ identical three-level atoms identically coupled to a single cavity mode. Each three-level system is represented by the transition frequencies
   $\omega_{10} = \omega_1-\omega_0 > 0$, $\omega_{21} = \omega_2-\omega_1 > 0$, $\omega_{20} = \omega_{21}+\omega_{10} > 0$ and by the oscillator strengths $f_{01} \geq 0$, $f_{12} \geq 0$, $f_{02} \geq 0$
   (see definition (\ref{f}) in the text).   \label{shapes}}
\end{figure}
 \begin{figure}
\begin{center}
\includegraphics[width=200pt]{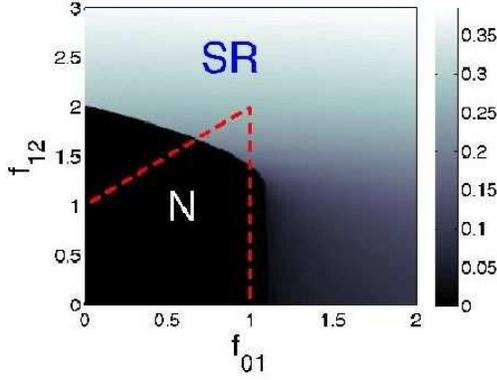}
\caption{\label{ladder} Results for the {\it ladder} configuration ($f_{02}=0$). Photonic order parameter $x = \alpha/\sqrt{N} = \langle a \rangle/\sqrt{N}$ in the ground state as a function of the oscillator strengths $f_{01}$ and $f_{12}$ for each three-level atom. The normal phase (N)  is characterized by $x = 0$ (black color). In the superradiant  phase (SR),  there is a spontaneous coherence  $x \neq 0$. Note that a discontinuous jump of $x$ denotes a first-order superradiant phase transition, otherwise the transition is of second order.
The area below the red-dashed  lines indicate the region described by the TRK inequalities $0 \leq f_{01} \leq 1$ and $0  \leq f_{10} +  f_{12}  \leq 1$ where $f_{10} = - f_{01}$. Parameters: $D = 3 \omega_{cav}$, $\omega_{10} = 0.1 \omega_{cav}$, $\omega_{21} = \omega_{cav}$.
The collective vacuum Rabi frequency $\Omega_{01}$ and $\Omega_{12}$ are obtained through the relationship in Eq. (\ref{D}).  
}
\end{center}
\end{figure}
\begin{figure}
\begin{center}
\includegraphics[width=200pt]{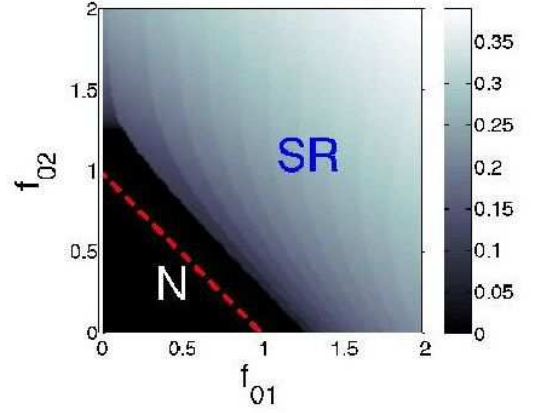}
\caption{\label{V} V-type configuration ($f_{12} =0$). Photonic order parameter  as a function of the oscillator strengths $f_{01}$ and $f_{02}$. 
The red-dashed  lines indicate the boundaries imposed by the TRK sum rule, namely $0 \leq f_{01} + f_{02} \leq 1$.
Parameters: $D = \omega_{cav}$, $\omega_{10} =  \omega_{cav}$, $\omega_{21} = 0.1 \omega_{cav}$.
In this configuration, the superradiant phase is always incompatible with the TRK boundaries.
}
\end{center}
\end{figure}
\begin{figure}
\begin{center}
\includegraphics[width=200pt]{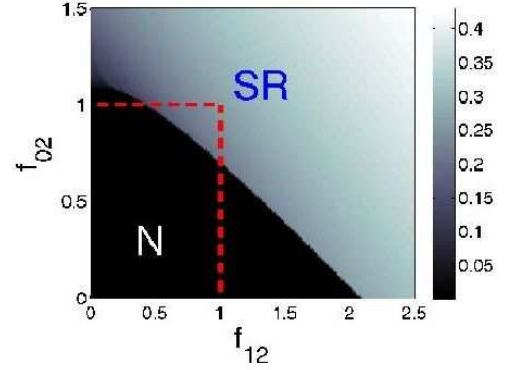}
\caption{\label{Lambda} Results for the {\it lambda} configuration ($f_{01} =0$). Photonic order parameter  versus $f_{12}$ and $f_{02}$. 
The area below the red-dashed  lines is compatible with the TRK inequalities, namely $0 \leq f_{12} \leq 1$
and  $0 \leq f_{02} \leq 1$.
Parameters: $D = 3  \omega_{cav}$, $\omega_{10} = 0.1 \omega_{cav}$, $\omega_{21} = 0.9 \omega_{cav}$.
As in the ladder case, the superradiant part of the diagram has an overlap with the region compatible with the TRK inequalities. 
}
\end{center} 
\end{figure}
As sketched in Fig. 1, let us consider $N$ identical three-level systems, whose states are $\left \{ \vert 0_k\rangle \hbox{,} \vert 1_k\rangle \hbox{,} \vert2_k\rangle \right \}\hbox{ , }( k=1,2,..,N) $ where $k$ is the atomic index. Each atom is assumed to be independent from the others and to interact identically with a single bosonic, photonic mode. The energies of the three levels are  $\hbar \omega_0< \hbar \omega_1< \hbar \omega_2$.

\begin{figure}
   \begin{center}
   \includegraphics[width=200pt]{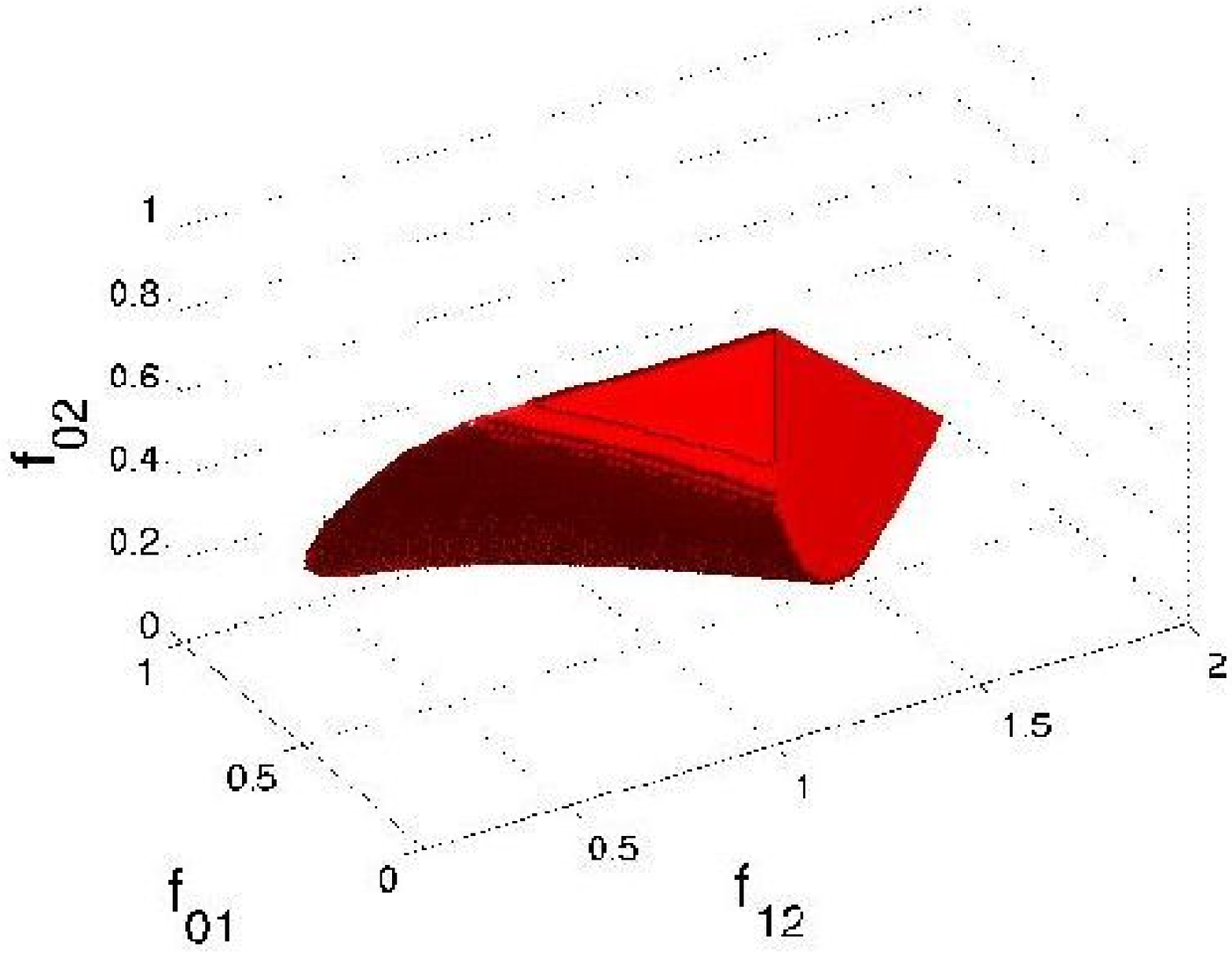} \\
     \includegraphics[width=180pt]{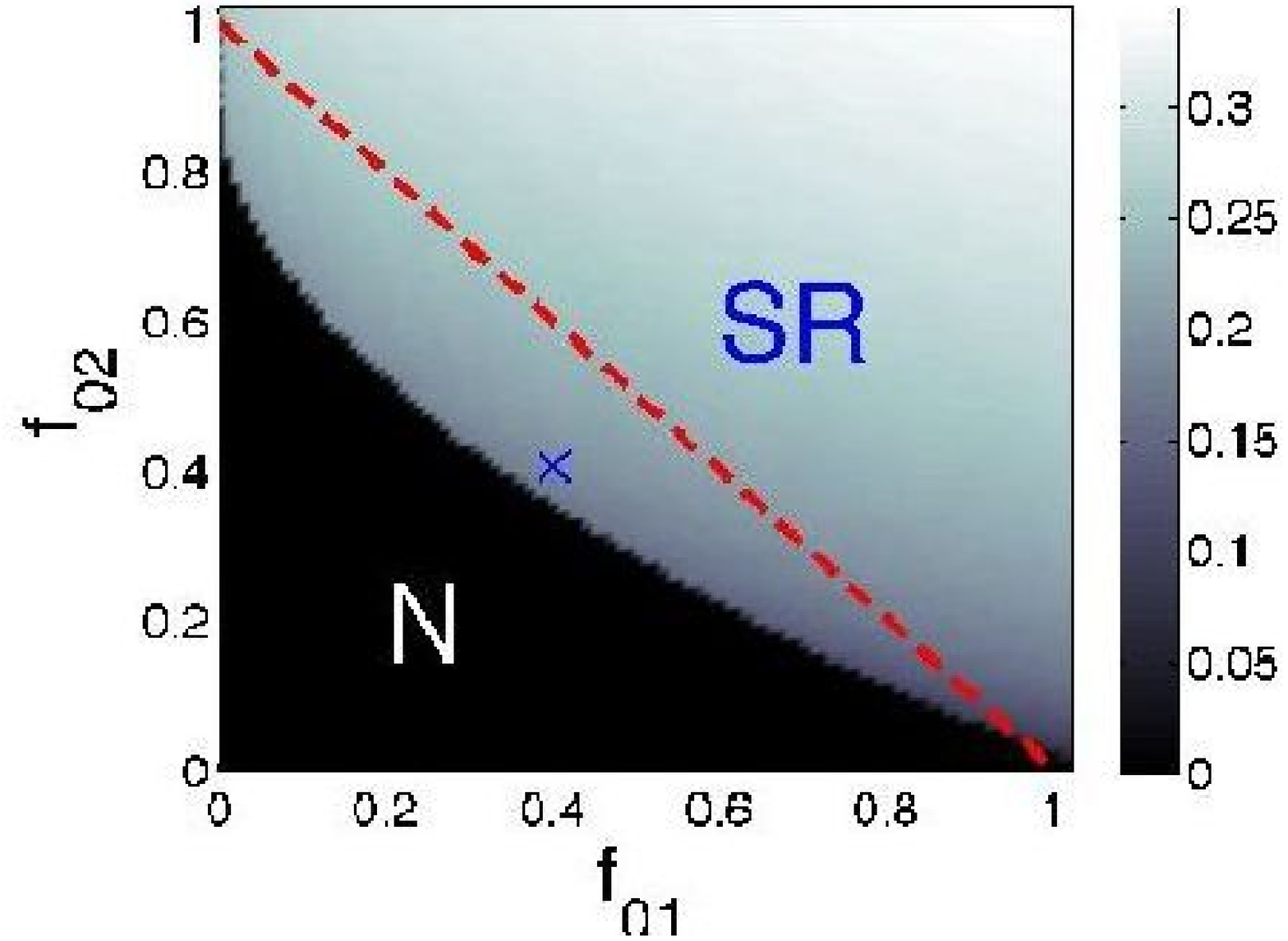} \\
    \includegraphics[width=170pt]{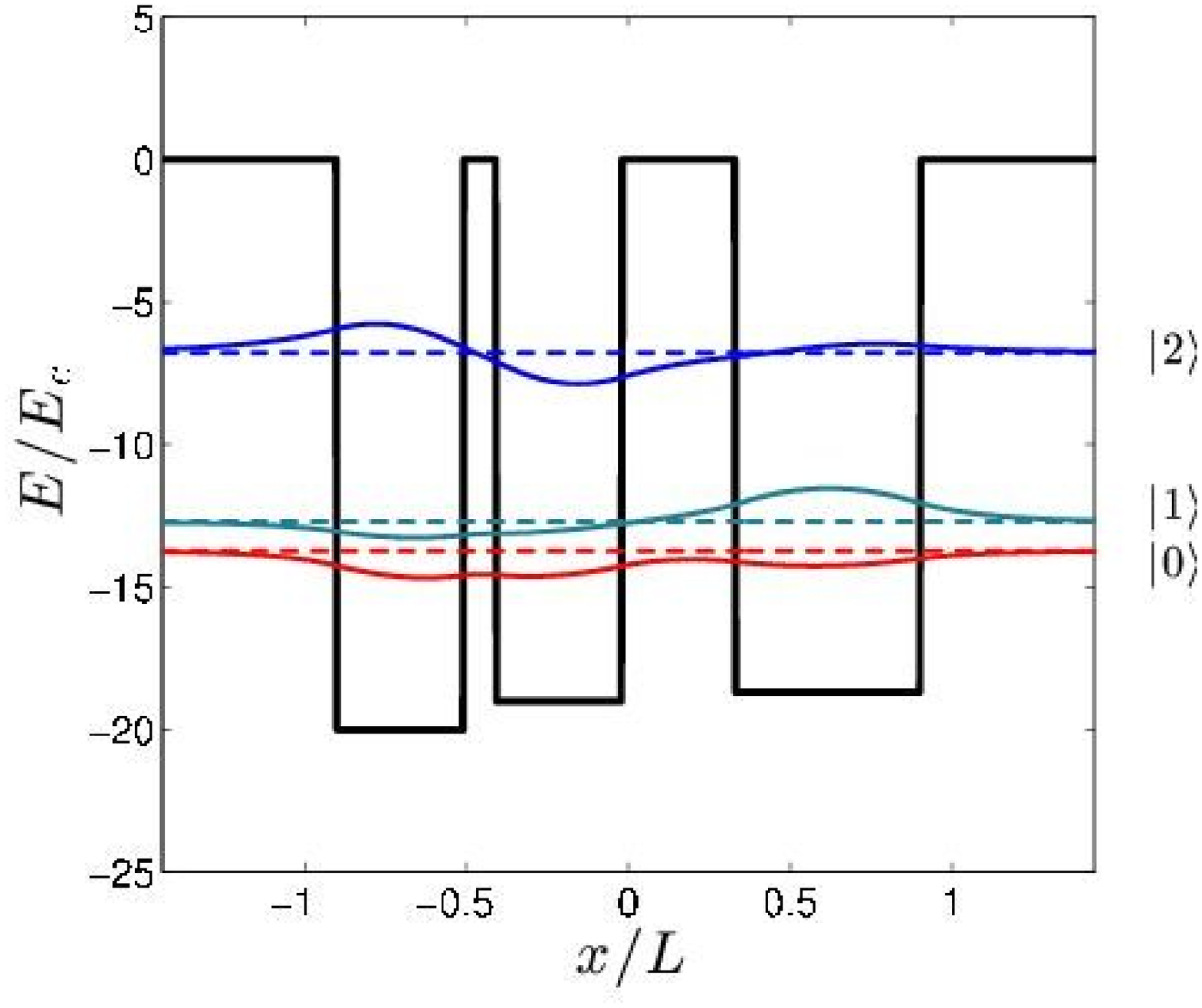}
   \end{center}
   \caption{ Top panel: the filled red volume in the $(f_{01},f_{02},f_{12})$ space represents the superradiant part of the diagram compatible with the TRK inequalities $0\leq f_{01} + f_{02} \leq 1$ and $0\leq f_{10} + f_{12} \leq 1$. Parameters: $D = 5 \omega_{cav}$, $\omega_{10} = 0.17  \omega_{cav}$, $\omega_{21} = \omega_{cav}$. Middle panel: the photonic order parameter $x$ as a function of $f_{01}$ and $f_{02}$ for a fixed value $f_{12} = 0.735$. The area below the red-dashed line corresponds to the region satisfying the TRK inequalities. 
  Bottom panel: the black line represents the spatial profile of an illustrative artificial one-dimensional potential with squared wells (as those realizable with semiconductor heterostructures)
 in units of the energy $E_c = \frac{\hbar^2}{2 m L^2}$ where the spatial coordinate is expressed in units of the length $L$. The horizontal lines depict the energies of the three bound states with their corresponding wavefunctions
 (other states are in the continuum). For this potential shape, one obtains the oscillator strengths $(f_{01} = 0.3995 , f_{02} = 0.4069,  f_{12}=0.735)$   and anharmonicity ratio $\omega_{10}/\omega_{21}= 0.1709 $  corresponding to the point depicted by
 the blue cross in the middle panel.    \label{SRVol}}
\end{figure}

By introducing the collective operators $
\hat{\Sigma}_{i j}=\sum_{k =1}^{N}  \vert i_{k} \rangle\langle j_{k} \vert $, the light-matter Hamiltonian reads 
\begin{equation}
\begin{array}{cc}
\label{H}
\mathcal{H}/\hbar= & \omega_{cav}  \hat{a}^{\dagger}\hat{a}+  \sum_{j=0}^2 \omega_j \hat{\Sigma}_{jj} + D(\hat{a}+\hat{a}^{\dagger})^2  \\
& +\sum_{\substack{i,j=0\\(i \neq j)}}^2 \Omega_{ij} \frac{1}{\sqrt{N}} (\hat{\Sigma}_{ij}+\hat{\Sigma}_{ji})(\hat{a}+\hat{a}^{\dagger}).
\end{array}
\end{equation}
The cavity mode is described by the frequency $\omega_{cav}$ and by its creation (annihilation) bosonic operator $a^{\dagger}$ ($a$). The coupling between the
cavity mode and the atomic $i \to j$ transition is quantified by the collective vacuum Rabi frequency $\Omega_{ij}$. $\Omega_{ij}/\sqrt{N}$ is the light-matter coupling per atom.
The term proportional to $D$ and quadratic in the photon operators is the so-called diamagnetic term.  It originates from the $\frac{(\mathbf{\hat{p}} -q \mathbf{\hat{A}})^2}{2m} $ form of
the non-relativistic electron-light interaction in the so-called $\mathbf{\hat{p}}\cdot \mathbf{\hat{A}}$ gauge, where $\mathbf{\hat{p}}$ is the electron momentum operator, $q$ the electron charge and $\mathbf{\hat{A}}$ is the  electromagnetic vector potential operator. As it can be deduced by the treatment in Ref. \cite{NatafNAT}, the value of the vacuum Rabi frequencies $\Omega_{ij}$
are linked to the diamagnetic term amplitude via the oscillator strengths as follows:
\begin{equation}
\label{D}
\frac{\Omega_{ij}^2}{\omega_{ji}} = f_{ij} D 
\end{equation} 
where $\omega_{ji} = \omega_{j}-\omega_i$ is the transition frequency and 
\begin{equation} \label{f} f_{ij} = \frac{2 m}{\hbar} (\omega_j -\omega_i) \vert d_{ij} \vert^2 
\end{equation}
 is the transition oscillator strength with $d_{ij}$ the corresponding atomic electric dipole  matrix element (along the direction of the photon mode polarization).
Indeed, as shown in Ref. \cite{NatafNAT} , $\hbar D = \frac{q^2}{2 m} n_{el} N \mathbf{A}^2_0$ where $q$ is the electron charge, $n_{el}$ the number of electrons per atom and  $\mathbf {\hat{A}} = {\mathbf A}_0 (a + a^{\dagger})$ is
the electromagnetic vector potential in the position where the atoms are assumed to be coupled (the spatial variation of the cavity field is neglected in the region occupied by the atoms, i.e., the electric dipole approximation has been considered). For a given collection of atoms and for a given cavity, $D$ is a constant depending on the density of atoms in the cavity volume via the factor $N \mathbf{A}^2_0$.
A related similar model can be obtained by considering a two-dimensional electron gas in a semiconductor quantum well heterostructure\cite{Ciuti2005,Anappara2,Anappara,Todorov} (replacing the bare electron mass with the semiconductor conduction band effective mass).

 The TRK oscillator strength reads 
 $\sum_{j} f_{i j} = 1$.  Note that by definition the oscillator strengths $f_{ij} > 0$ if $\omega_{j} > \omega_i$, while it is negative if $\omega_{j} < \omega_i$.
  Hence, for a transition from the ground level $\vert 0\rangle$ to the first excited level $\vert 1 \rangle$, we have always
  $0 \leq f_{01} \leq 1 $ as $f_{0j} \geq 0$.
For two-level system models, the presence of the diamagnetic term and the constraint $0 \leq f_{01} \leq 1$ is sufficient to prevent the superradiant phase transition (no-go theorem)\cite{pol1,NatafNAT}.
Here, we will analyze what happens with three-level systems.
By applying the method detailed in Ref. \cite{Molmer} , we can express the generalized collective transition operators by using  a multilevel Holstein-Primakoff transformation in terms of bosonic annihilation (creation) operators
$\hat{c}_k$ ( $\hat{c}^{\dagger}_k$).
When we use this procedure for the three-level case, we have to choose a state of reference that we will call $r$.  The multilevel boson mapping reads:
\begin{align}
& \hat{\Sigma}_{kk}=\hat{c}^{\dagger}_k\hat{c}_k\phantom{...}(k\neq r) \hspace{0.5cm} \hbox{,} \hspace{0.5cm}\hat{\Sigma}_{rr}=N-\sum_{k\neq r} \hat{c}^{\dagger}_k\hat{c}_k \nonumber \\
& \hat{\Sigma}_{ir}=\hat{c}^{\dagger}_i\sqrt{N-\sum_{k\neq r} \hat{c}^{\dagger}_k\hat{c}_k} \hspace{0.5cm} \hbox{,} \hspace{0.5cm} \hat{\Sigma}_{ij}=\hat{c}^{\dagger}_i\hat{c}_j\phantom{...}(i,j\neq r) \,.\end{align}
The collective transition operators $\hat{\Sigma}_{ij}$ defined above do not commute and are such that $\left[ \hat{\Sigma}_{ij},\hat{\Sigma}_{kl}\right]=\delta_{jk}\hat{\Sigma}_{il}-\delta_{il}\hat{\Sigma}_{kj}$.
The Hamiltonian can be rewritten as
\begin{align}
\mathcal{H}/\hbar = & \omega_{cav} \hat{a}^{\dagger}\hat{a}+ \sum_{k\neq r}(\omega_k-\omega_r) \hat{c}_k^{\dagger}\hat{c}_k+ \omega_r N \nonumber \\ 
&+ \Bigg[\sum_{k \neq r} \frac{\Omega_{kr}}{\sqrt{N}} \left(\hat{c}_k^{\dagger}\sqrt{N-\sum_l \hat{c}^{\dagger}_l\hat{c}_l}  +\sqrt{N-\sum_l \hat{c}^{\dagger}_l\hat{c}_l  \phantom{..}}  \hat{c}_k \right) \nonumber \\ 
& + \sum_{\substack{i>j\\(i,j \neq r)}}\frac{\Omega_{ij}}{\sqrt{N}} (\hat{c}_i^{\dagger}\hat{c}_j + \hat{c}_j^{\dagger}\hat{c}_i) \Bigg](\hat{a}+\hat{a}^{\dagger}) + D(\hat{a}+\hat{a}^{\dagger})^2  \label{general}
\end{align}

In the following, we choose $r = 1$ as reference state (choosing another reference state leads to the same results).
In order to determine the phase diagram in the thermodynamical limit, we can use the mean-field approach as in Refs. \cite{Brandes,Hayn}, by  the replacement $\langle a \rangle=\langle a^{\dagger}\rangle=\alpha$ , $\langle c_j \rangle=\langle c_j^{\dagger}\rangle=\beta_j$  with $j \in \{0,2 \} $. 
It can be shown that due to the form of Eq. (\ref{H}) the solutions for $\alpha$ and $\beta_j$ are necessarily real numbers.
Therefore, we obtain the mean-field expression of the ground state energy (\ref{general}) 
\begin{align}
E_G/\hbar= & (\omega_{cav}+4D)\alpha^2+\omega_{01}\beta_0^2+\omega_{21}\beta_2^2+\omega_1 N \nonumber \\
 & +4\alpha\left[(\Omega_{10}\beta_0+\Omega_{21}\beta_2)\sqrt{N-\beta_0^2-\beta_2^2}+\Omega_{20}\beta_0\beta_2\right],
\end{align}
whose global minimization will give the values of $\alpha$ (photonic coherence), $\beta_0$ and $\beta_2$ (collective atomic coherences) in the ground state.
A photonic order parameter $\alpha = 0$ implies that the system in the Normal (N) phase. If $\alpha \neq 0$, a SuperRadiant (SR) phase occurs\cite{Brandes}.
 In order to minimize the ground state energy in the thermodynamical limit ($N \to \infty $) it is convenient to introduce the rescaled quantities $x=\frac{\alpha}{\sqrt{N}}$ , $y=\frac{\beta_0}{\sqrt{N}}$ and $z=\frac{\beta_2}{\sqrt{N}}$. The ground state energy can therefore be rewritten as
 \begin{align}
\frac{E_G/\hbar}{N}= & (\omega_{cav}+4D)x^2+\omega_{01}y^2+\omega_{21}z^2+\omega_1\nonumber \\
 & +4x\left[(\Omega_{10}y+\Omega_{21}z)\sqrt{1-y^2-z^2}+\Omega_{20}yz\right] \label{EGxyz} \, .
\end{align}
As $\frac{\partial^2 E_G}{\partial x^2} = 2 (\omega_{cav} + 4 D) > 0 $, the minimization with respect to $x$ is therefore trivial, giving $x$ as a function of $y$ and $z$. In order to minimize the ground state energy, we therefore need to minimize with respect to $y$ and $z$ the following energy function (obtained from Eq. (\ref{EGxyz})):
 
 \begin{align}
\mathcal{E}_G= & \omega_{01}y^2+\omega_{21}z^2+\omega_1\nonumber \\
 & -\frac{4}{\omega_{cav}+4D}\left[(\Omega_{10}y+\Omega_{21}z)\sqrt{1-y^2-z^2}+\Omega_{20}yz\right]^2\label{EGyz}\,.
\end{align}
Note that $\beta_0^2+\beta_2^2\leq N$  (which is equivalent to $y^2+z^2\leq 1$) must be fulfilled within the Holstein-Primakoff framework. Hence, the global minimum of $\mathcal{E}_G$ has to be looked for in a disk with unity radius centered in the origin of the $0yz$ plane, which can be done with a straightforward numerical calculation.

We start by considering the {\it ladder} configuration ($f_{02} = 0$) for the three-level systems. In Fig. \ref{ladder}, we have plotted results for the photonic coherence $ x = \alpha/\sqrt{N} =  \langle a \rangle/\sqrt{N}$ as a function of the oscillator strengths $f_{01}$ and $f_{12}$ of the ladder coupling
for $ D = 3 \omega_{cav}$, $\omega_{21} = \omega_{cav}$ and $\omega_{10}/\omega_{21} = 0.1$. We point out that these parameters correspond to a strong anharmonicity with the photon mode in resonance with the excited transition. The value of the collective vacuum Rabi frequencies is given by Eq. (\ref{D}). The normal phase occurs when $\alpha = 0$ (black region). The results in Fig. \ref{ladder} show that in the considered system a superradiant phase ($\alpha \neq 0$) do occur. The phase transition boundary can be of both of first and second order. The first-order phase transition occurs when $\alpha$ has a discontinuous jump from $0$ to a finite value (upper part of the frontier).  The first-order transition boundary is due to the excited transition $1\to 2$   
as in the case recently studied by Hayn {\it et al.}\cite{Hayn}: at the transition there is a macroscopic occupation of the intermediate state $\vert 1 \rangle$. The second-order phase transition occurs when $\alpha$ is not discontinuous, but its gradient is.  
The important point to consider here is the fact that due to the TRK sum rule the oscillator strengths are subject to constraints.
For the ladder configuration, we have $0 \leq f_{01} \leq 1$ and $0 \leq f_{10} + f_{12} \leq 1$, where $f_{10} = - f_{01} \leq 0$.
In Fig. \ref{ladder} the TRK boundaries imposed by such inequalities have been indicated: the area compatible with the TRK sum rule is below the red-dashed line. It is therefore apparent that there is an overlap between the superradiant part of the diagram and the region compatible with the TRK sum rule.  Note that the relevant superradiant phase region compatible with the TRK inequalities contains a first-order transition boundary.

In Fig. \ref{V}, we consider instead the V-type configuration, where $f_{12} = 0$, i.e., there is no coupling between the excited states.
In such a configuration, the photonic order parameter $x = \alpha/\sqrt{N} $ is plotted as a function of $f_{01}$ and  $f_{02}$ (parameters in the caption). 
The TRK sum rule imposes the inequality $0 \leq f_{01}+f_{02} \leq 1$, again indicated by the red-dashed line. In such a V-type configuration, there is a superradiant phase boundary, which is always of second order. In fact, the transition between excited states is by definition inactive in such a configuration. Importantly, we notice that here there is no overlap between the superradiant part of the diagram and the area compatible with the TRK sum rule. Indeed, the TRK sum rule always prevents the superradiant phase transition in the V-type configuration as in the two-level case\cite{NatafNAT}.

In Fig. \ref{Lambda}, we show results for the {\it lambda} configuration ($f_{01} = 0$): the photonic coherence is shown versus $f_{12}$ and $f_{02}$ (parameters in the caption). Here, the TRK inequalities $0 \leq f_{12} \leq 1$ and
 $0 \leq f_{02} \leq 1$, whose boundaries are once again delimited by the red-dashed line. As in the ladder case, in the lambda configuration there is a superradiant phase in the region compatible with the TRK inequalities (are below the red-dashed line) with the transition boundaries being of the first order.  
  
So far, we have shown the simplest three-level configurations (ladder, V-type and lambda). In Fig. \ref{SRVol}, we show results  for a generic three level system where all the three oscillator strengths are finite (detailed parameters in the caption). The superradiant part satisfying the TRK inequalities is depicted by the red filled volume in the $(f_{01},f_{02},f_{12})$ space (top panel). The middle panel shows the photonic order parameter  on a planar section of such oscillator strength three-dimensional space. We point out that in the generic three-level case the interesting superradiant part compatible with the TRK inequalities is widened due to the additional freedom associated to the third oscillator strengths . In the bottom panel of Fig. \ref{SRVol}, we give an illustrative example of a spatial artificial potential providing a three-level system with oscillator strengths $(f_{01},f_{02},f_{12})$ and anharmonic spectrum corresponding to the cross in the middle panel, for which a superradiant transition is possible while satisfying the TRK constraints. As a perspective, it will be interesting in the future to address more complex multilevel structures and to investigate also the role of direct Coulomb interactions\cite{Keeling}.

In conclusion, we have determined the phase diagram for a model system consisting of $N$ three-level systems coupled to a single photonic boson mode, by including the diamagnetic contribution in the light-matter coupling.
We have demonstrated that in the considered model system superradiant phase transitions can occur while preserving the TRK inequalities, in stark contrast to the case of two-level systems. We have found that the transition between excited levels  have a key role in such superradiant phase transition in contrast to what assumed in Ref. \cite{Viehmann}.   Our results show that the physics of superradiant phase transitions with multilevel systems might be achievable in a broad range of physical systems, especially those where it is possible to engineer the spectra and oscillator strengths. 
We would like to thank T. Brandes, I. Carusotto, C. Emary and J. Keeling for stimulating discussions. C. C. is member of {\it Institut Universitaire de France} (IUF). We acknowledge support from the ANR project QPOL.

\end{document}